\documentclass[prb,twocolumn]{revtex4}
\usepackage{epsfig}
\usepackage{times}
\begin{document}
\author{Dietrich Dehlinger, M. W. Mitchell}
\affiliation{~\\ Physics Department, Reed College 3203 SE Woodstock Blvd. 
\\
Portland, OR 97202}
\newcommand{\DateWritten}{\today}
\newcommand{\PutInDate}[1]{\hfill {\normalsize \rm #1}}
\email{morgan.mitchell@reed.edu}   
\title{Entangled photon apparatus for the undergraduate laboratory}
\newcommand{\FigWidth}{3in}
\begin{abstract} 
   
We present detailed instructions for constructing and operating an
apparatus to produce and detect polarization-entangled photons.  The 
source operates by type-I spontaneous parametric downconversion in a 
two-crystal geometry.  Photons are detected in coincidence by 
single-photon counting modules and show strong angular and 
polarization correlations.  We observe more than 100 
entangled photon pairs per second.  A test of a Bell 
inequality can be performed in an afternoon.
\end{abstract}
\maketitle
\newcommand{\be}{\begin{equation}}
\newcommand{\ee}{\end{equation}}
\newcommand{\bea}{\begin{eqnarray}}
\newcommand{\eea}{\end{eqnarray}}
\newcommand{\ket}[1]{\left|#1\right>}
\newcommand{\bra}[1]{\left<#1\right|}
\newcommand{\bk}{{\bf k}}
\newcommand{\etal}{{\em et al.}} 
\newcommand{\degree}{$^{\circ}$}

\section{Introduction} 
Entanglement of particles is ubiquitous in quantum mechanical systems
and is arguably the aspect of quantum theory most at odds with 
classical intuitions 
.
Einstein, 
Podolsky and Rosen first drew attention to the possibility
of non-local effects involving entangled particles.\cite{Einstein1935}  
In the mid-Sixties it was realized that the nonlocality of 
nature was a testable hypothesis  and subsequent 
experiments confirmed the quantum predictions.  More recently, much 
effort has gone into exploiting the odd nature of entangled particles.\cite{Bell1964}  
Applications include secure  
cryptography , transmission of 
two bits of information with a single photon and  
``teleportation'' of a quantum state (by erasing the state of a system 
and then re-creating the state in a distant 
system).\cite{Ekert1991,Naik2000,Jennewein2000,Bennett1992,Bennett1993} 
Perhaps the most exciting possible application is in computing.  A 
``quantum computer'' which used entangled particles as data registers 
would be capable of performing calculations faster than any classical
computer. 
\cite{Feynman1982,Deutsch1992,Shor1997,
Nielsen2000}

Here we describe how to produce and detect entangled photons
using equipment and techniques suitable for an undergraduate 
laboratory.  This is possible due to recent advances in diode laser 
technology and new techniques for generation of photon pairs 
.\cite{Kwiat1995,Kwiat1999}  The total cost for the apparatus is
approximately 15,000 USD.



\begin{figure}[h]
\centerline{\epsfig{width=\FigWidth,figure=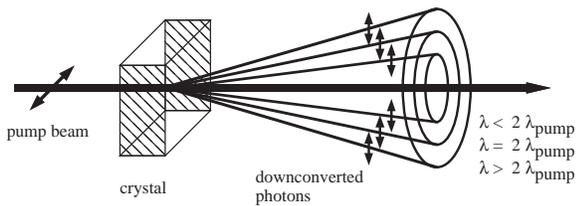}}
\label{fig:OneCrystal}
\renewcommand{\baselinestretch}{1}
\caption{Type I spontaneous parametric downconversion.  Photons from a 
pump beam split into pairs of photons.}
\renewcommand{\baselinestretch}{2}
\end{figure}

\begin{figure}[h]
\centerline{\epsfig{width=\FigWidth,figure=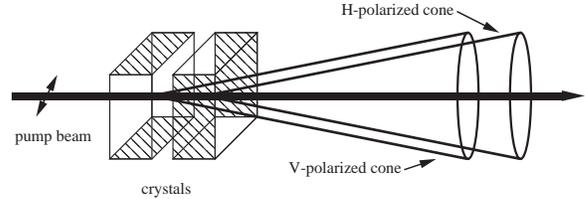}}
\label{fig:TwoCrystal} 
\renewcommand{\baselinestretch}{1}
\caption{Two-crystal downconversion source.}
\renewcommand{\baselinestretch}{2}
\end{figure}

\section{Overview}

In the process of spontaneous parametric downconversion (SPD)
 a single pump photon spontaneously
splits into ``signal'' and ``idler'' photons inside a nonlinear
crystal.\cite{Hong1985,Hariharan1996}  Because the two downconverted photons come from a single parent photon
they have definite combined properties: their total energy and
momentum (inside the crystal) must agree with the parent energy and
momentum.  They are also produced at very nearly the same time.
The individual photons' properties are free to vary, however, and
SPD produces a spectrum of both signal and idler wavelengths centered
around twice the parent photon wavelength.  To satisfy
phase matching requirements, which reflect the dispersion and 
birefringence of the crystal, these different wavelengths emerge in 
different directions and create a conical rainbow of emission as 
illustrated in Figure 
1
.  

By appropriate angular placement of the detectors, we
select ``degenerate'' daughter photons,
those which have the same wavelength $\lambda_{\rm S} =\lambda_{\rm I}
= 2 \lambda_{\rm pump}$.  Our crystals are cut for Type I 
downconversion, in which the
signal and idler photons have the same polarization, which is opposite
to that of the pump photon. \cite{Boyd1992}  A given crystal can only
support Type I downconversion of one pump polarization, the other
polarization simply passes through the transparent crystal.
Our source uses two identical crystals, with one rotated 90\degree~ 
from the other about the beam propagation direction, as shown in
Figure 
2.
In this arrangement
each crystal can support downconversion of one pump polariazation.
A 45\degree~ polarized pump photon can downconvert in either crystal, 
producing a polarization-entangled pair of photons. \cite{Kwiat1999}


\begin{figure}[h]
\centerline{\epsfig{width=\FigWidth,figure=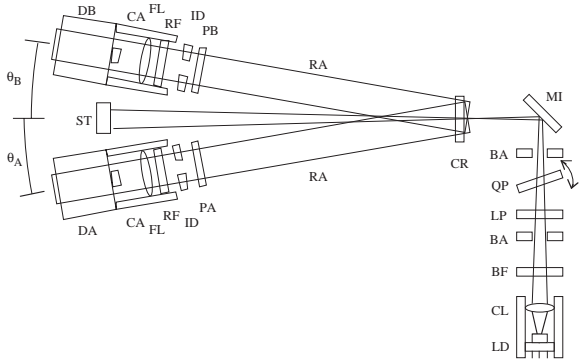}}
\label{fig:Setup} 
\renewcommand{\baselinestretch}{1}
\caption{Schematic diagram of experimental setup, not to scale.  
Symbols:  LD Laser Diode, CL 
Collimating Lens, BF Blue Filter, BA Beam Aperture, LP Laser Polarizer, QP
Quartz Plate, MI Mirror, CR Downconversion Crystals, RA Rail, PA 
Polarizer A, PB Polarizer B, ID Iris Diaphragm, RF Red Filter,
FL Focusing Lens, CA Cage Assembly, DA Detector A, DB Detector B, ST 
Beam Stop.}
\renewcommand{\baselinestretch}{2}
\end{figure}

Figure 
3
shows a schematic of the experimental setup.  A 
5 mW, 405 nm, InGaN laser diode (Nichia Model NLHV500C) is the source 
of pump photons.  The diode is driven at its specified operating 
current with a precision constant current source capable of supplying 
the 5V operating voltage of the InGaN laser.  
The beam is focused with an aspheric lens 
mounted in a collimating tube, which is held in a vee-mount 
glued to a mirror mount.  The beam focus is adjusted to 
produce a beam diameter of about 1 mm at 1 m from the laser.

The laser beam passes a BG12 colored glass filter, a beam aperture 
consisting of a sheet of aluminum with small hole drilled in 
it, a glass linear polarizer on a rotating mount, a 0.5 mm thick 
quartz plate mounted to rotate about the vertical (the crystal's fast 
axis) and a second beam aperture before reaching the nonlinear crystals.  
The apertures and filter act to remove unwanted light and the polarizer 
and quartz plate set the laser polarization.

The nonlinear crystals are two beta barium borate (BBO) crystals, 5 mm 
$\times$ 5 mm $\times$ 0.1 mm thick, cut with their crystal axes at 
29\degree~ from normal to the large face.  
The crystals are mounted face-to-face with one crystal 
rotated by 90\degree~ about the normal to the large face. 
The crystal 
holder is held in a 
mirror mount and aligned to retroreflect the laser beam.  A beam stop 
blocks the laser beam after it passes alongside the detectors.

The downconverted photons produced in the BBO crystals travel about
1 meter before passing an adjustable 1 inch diameter iris diaphragm, a
plastic film near-infrared polarizer on a rotatable mount, a 1 inch
diameter RG780 colored glass filter, and are focused by a 75mm focal
length lens onto the detector surface.  
The RG780 is a long-pass filter with a 50\% transmission at
780 nm.  For a coincidence detection event to occur, both signal and 
idler photons, which are roughly equally spaced in wavelength about
810 nm, must pass RG780 filters.  Thus a wavelength band (in the 
signal photon) of about 780--840 nm can lead to coincidence 
detection.  

The detectors (Perkin-Elmer Optoelectronics model SPCM-AQR-13) are 
silicon avalanche photodiodes run in Geiger mode, called single-photon 
counting modules (SPCMs). \cite{PerkinElmer,Ghioni1996}  

To aid in focusing, each detector was mounted on standard optical 
mounting posts and a 30 mm ``cage'' assembly was attached to hold the 
lens and RG780 filter.  These are described in section 
\ref{sec:Mechanics}.  The lens is held in an X-Y translator for fine 
adjustment of the lens position.

Along each detection path, the detector, lens, filter, polarizer and iris
are mounted on an aluminum rail which pivots about an optical post
directly below the crystal.  This arrangement allows adjustment of the
angular position of the detectors without losing focus.  Although we
performed our experiments on an optical table, we note that this is
probably unncecessary.  Interferometric stability is not required and
the rails maintain the necessary alignment.  We expect the experiment
could be performed on an optical breadboard or other flat surface.




\begin{figure}[h]
\centerline{\epsfig{width=\FigWidth,figure=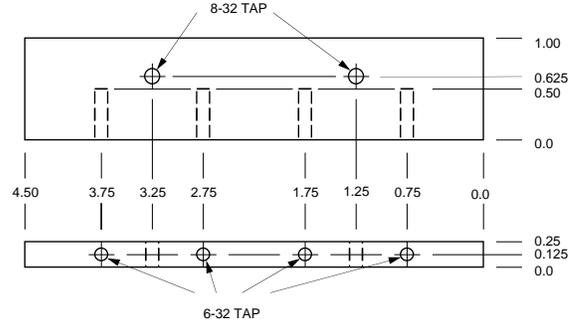}}
\label{fig:BaseAdapter}
\renewcommand{\baselinestretch}{1}
\caption{SPCM base adapter plate.  Units: inches, all tolerences are
$\pm$ 0.01 inch unless otherwise specified.}
\renewcommand{\baselinestretch}{2}
\end{figure}

\begin{figure}[h]
\centerline{\epsfig{width=\FigWidth,figure=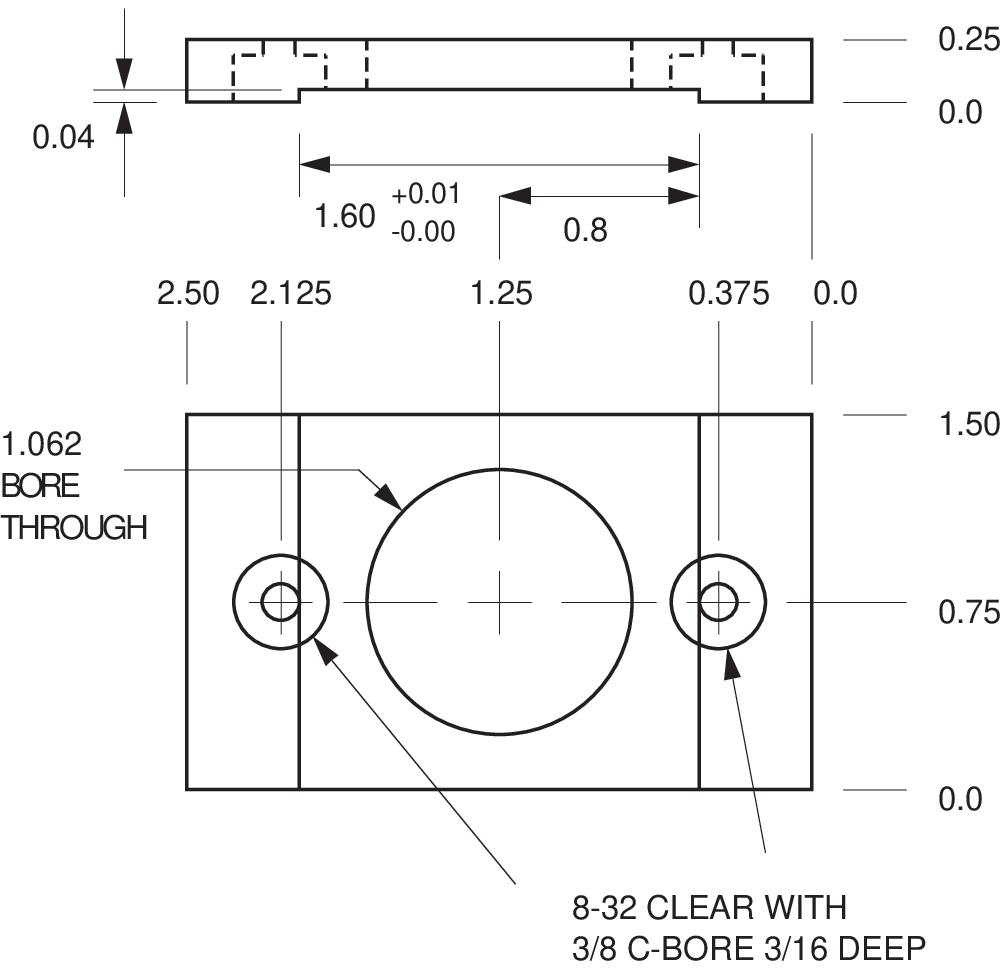}}
\label{fig:FrontAdapter}
\renewcommand{\baselinestretch}{1}
\caption{SPCM front adapter plate.  Units: inches, all tolerences are
$\pm$ 0.01 inch unless otherwise specified.}
\renewcommand{\baselinestretch}{2}
\end{figure}

\begin{figure}[h]
\centerline{\epsfig{width=\FigWidth,figure=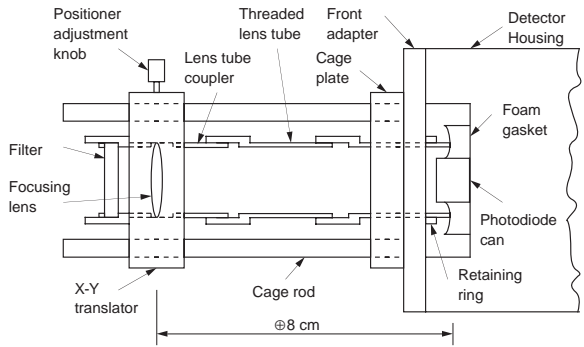}}
\label{fig:CageAssembly}
\renewcommand{\baselinestretch}{1}
\caption{Schematic diagram of detector-mounted cage assembly, 
side view.}
\renewcommand{\baselinestretch}{2}
\end{figure}

\section{mechanics}
\label{sec:Mechanics}

The first version of the detection setup used independent
table-mounted components to hold the filters and position the focusing
lenses.  This early setup was quick to assemble from common components
such as translation stages and lens holders, but was difficult to
align and required near-complete darkness.  For greater ease of use
we developed the mechanical system described below.  

The SPCMs, as furnished by the manufacturer, do not directly interface
to standard optical hardware.  We built two simple adapter plates,
shown in Figures 
4
and 
5.
The base adapater allows the detector to be mounted on standard
optical posts, while the front adapter allows a cage assembly to be
attached to the detector, as shown in Figure
6.  
The cage assembly consists of four rods,
each 4 inches long, held at the detector end by a 30 mm cage plate and
at the other end by a 30 mm X-Y translator which holds the focusing
lens and filter.  The cage plate attaches to the front adapter by a
threaded tube and a retaining ring.  This same tube presses against a
gasket of black closed-cell foam to make a light-tight seal.  Other
threaded tubes link the X-Y translator to the cage plate.  This
arrangement has two principal advantages.  First, the path from filter
to detector is entirely enclosed, greatly reducing sensitivity to
stray light and protecting the detector from mechanical damage. 
Second, precise positioning of the focusing lens is easily
accomplished by sliding the X-Y translator along the rails and
adjusting the X-Y translation knobs.  Because the lens sits about 1m
away from the crystals, the image of the crystals on the detector
moves by about 75 $\mu$m for every 1 mm displacement of the lens. 
This is handy as the detector area is only 180 $\mu$m in diameter.


\begin{figure}[h]
\centerline{\epsfig{width=\FigWidth,figure=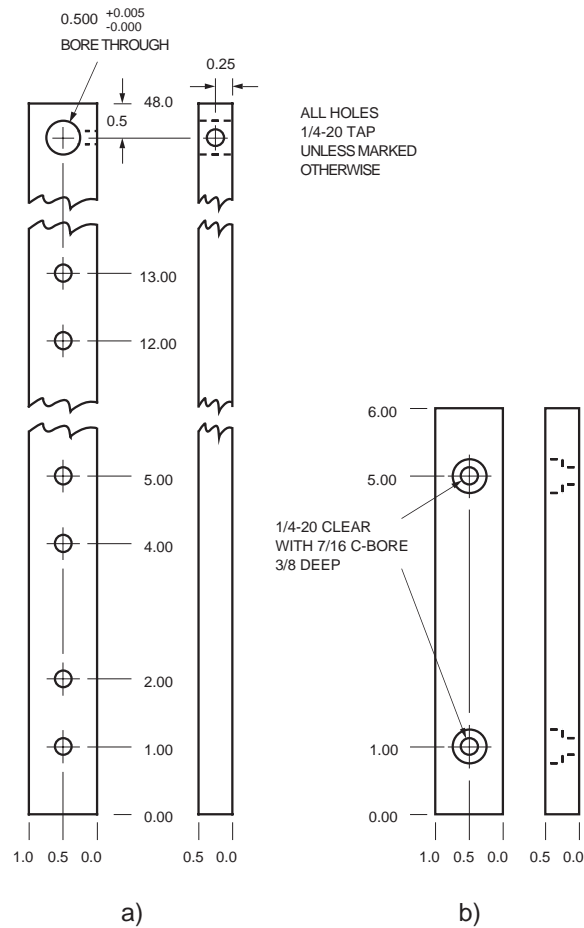}}
\label{fig:DetectorRail}
\renewcommand{\baselinestretch}{1}
\caption{Detector rail.  a) rail b) under-rail spacer.  All parts are 
aluminum.  Units: inches, all tolerences are $\pm$ 0.01 inch unless 
otherwise specified.}
\renewcommand{\baselinestretch}{2}
\end{figure}

Each detector and cage assembly, as well as the associated iris and
polarizer, sit upon a detector rail as shown in Figure
3.
We constructed our rails from half-inch thick
aluminum bars, as shown in Figure 
7.
Four post
holders are mounted directly to the rail to hold the detector, iris,
and linear polarizer.  A half-inch diameter hole at the opposite end
from the detector fits over a standard (0.499 inch diameter) optical
post which serves as a pivot for the rail.  The second rail pivots
about the same post, sitting upon the first rail.  A half-inch high
spacer beneath the second rail on the detector end allows the rail to
sit flat on the table.


\begin{figure}[h]
\centerline{\epsfig{width=\FigWidth,figure=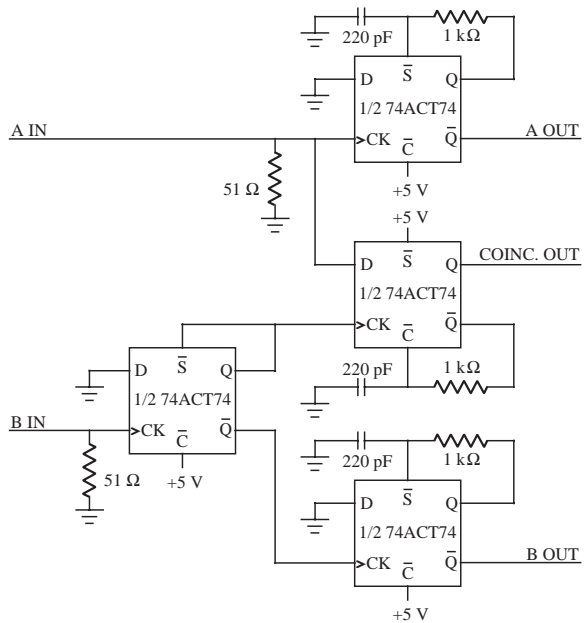}}
\label{fig:CoincCircuit}
\renewcommand{\baselinestretch}{1}
\caption{Schematic diagram of coincidence detection circuit.  See text 
for details.}
\renewcommand{\baselinestretch}{2}
\end{figure}

\section{electronics}

Detection of a photon by an SPCM produces a short (about 25 ns) TTL
pulse.  To identify photons from SPD within
the background of fluorescence photons, we record coincident
detections.  Our first coincidence detector used a NIM standard
time-to-amplitude converter and a multi-channel analyzer (MCA).  This
setup required converting the TTL pulses to NIM pulses and also
required a computer to record data from the MCA.  A simpler and far
less expensive alternative is to build a coincidence circuit from fast
logic chips.  A schematic for such a circuit is shown in Figure
8.

The circuit is built from two 74ACT74 dual D-type positive
edge-triggered flip-flops, or four D-type flip-flops in all.
Starting at the lower left of the diagram and working clockwise, the
first flip-flop delays the B pulse by the time required to clock and 
reset, 6.5 ns to 19.5 ns by the manufacturer's specifications.  The 
next flip-flop produces a
stretched pulse at A OUT. The next performs the coincidence detection:
If A IN is already high when the delayed B pulse arrives, a stretched
pulse is produced at COINC OUT. The coincidence window is the duration
of the A IN pulse.  The final flip-flop stretches the B pulse.  The 
inputs are 50 $\Omega$ terminated.  The
outputs pusles are all TTL compatible and about 250 ns long.  Longer pulses
can be produced by increasing the values of the capacitors.  The 
output pulses are fed to the counter-timer inputs of a PC-based
data acquisition board.  

The coincidence circuit is simple to build, but as with any fast
electronics, proper construction technique is necessary for good
performance.  It is helpful to have a fast oscilloscope when debugging
this circuit.

\section{alignment}

The apparatus is not difficult to align if done in the correct order. 
All optical elements should be set to the same height above the table. 
With the crystals, polarizers and RG780 filters removed from the setup,
the remaining rail-mounted components (detectors, lenses, and irises)
can be aligned as follows.  The detector is viewed from the position
the crystals will occupy, through the open iris and lens.  The detector's
active area is a circular black spot at the center of a gold square. 
By aligning and focusing the lens, the detector's active area can be
made to fill the field of view.  The iris should be centered on
this image.


\begin{figure}
\centerline{\epsfig{width=\FigWidth,figure=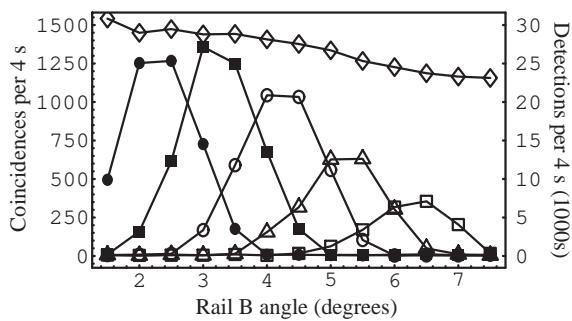}}
\label{fig:CoincVsAngle} 
\renewcommand{\baselinestretch}{1} 
\caption{Coincidence and singles detection rates as a function of
detector positions.  Solid circles, solid squares, hollow circles,
hollow triangles and hollow squares show coincidence rates for
$\theta_{A}$ = 2\degree~, 3\degree~, 4\degree~, 5\degree~, and
6\degree~, respectively, on left scale.  Diamonds show channel B
detection rate on right scale.  } 
\renewcommand{\baselinestretch}{2}
\end{figure}

Once this rough alignment is performed, the crystals, polarizers and
RG780 filters can be put in place.  The laser is aligned to pass
through the crystals' center.  The rails should be positioned on
opposite sides of the laser beam, each about three degrees away from
the beam.  With the room darkened, fine adjustments in lens position
can be made to maximize the singles count rate at each detector. 
Finally, the angles of the rails should be adjusted to maximize the
coincidence detection rate.  Following this procedure we find
coincidence rates greater than 300 cps, as shown in Figure
9.

To show that the photons are polarization-entangled, we measured
coincidence rates as a funciton of the two polarizer orientations. 
The polarization correlations were strong enough to demonstrate a
violation of a Bell inequality.  We used the the Clauser, Horne,
Shimony and Holt version of the Bell inequality $|S| \le 2,$ where $S$
is a measure of the polarization correlation involving sixteen
coincidence measurements.  As described in the accompanying
paper, we found $S = 2.307 \pm 0.035,$ a clear
violation.\cite{Dehlinger2002B}  The total acquisition time in this experiment was 240
seconds and a fit to the coincidence data indicates more than 100
polarization-entangled photons per second.  These data were taken with 
the irises fully opened.  Closing down the irises will reduce the 
count rate, but may improve the purity of the detected entangled 
state.

\section{miscellaneous}

A darkened room is necessary for the experiment but complete darkness
is neither necessary nor desirable.  In our setup, black velvet
curtains surrounding the work area block out sunlight and room lights. 
We find a dimmed but visible computer monitor adjacent to the optical
table gives negligible background coincidence counts.  We use green
LEDs behind BG-18 colored glass filters to illuminate the table; the
green light is blocked by the RG780 filters.

\section{acknowledgements}
We thank Paul Kwiat for inspiration and helpful 
discussions.  This work was supported by Reed College and 
grant number DUE-0088605 from the National Science Foundation.

\appendix
\section{Suppliers and equipment notes}
\label{app:suppliers}

\newcommand{\supplier}[1]{\noindent\newline #1 }

Below we list components of the apparatus which are unique or 
difficult to find, as well as suppliers for these components.  
These are also the most costly parts of the apparatus.  Remaining
components not described below include standard optical hardware
such as mirror mounts, electronics to build the coincidence 
detector, and a means to record the detection rates.

\supplier{
Edmund Industrial Optics\\
101 East Gloucester Pike\\
Barrington, NJ 08007-1380\\
(800) 363-1992\\
http://www.edmundoptics.com\\
\\
}
25.4 mm diameter unmounted linear glass polarizing filters.
1''$\times$1'' Polaroid near-IR linear polarizing film $\times$2.
1'' diameter RG-780 colored glass longpass filter $\times$2.
1'' diameter BG-12 blue-violet colored glass bandpass filter.
1'' diameter BG-18 blue-green colored glass bandpass filter.
Total cost:  \$255.

\supplier{
ILX Lightwave\\
P.O. Box 6310\\
Bozeman, MT \\
(800) 459-9459\\
http://www.ilxlightwave.com\\
\\
}
LDX-3412 Precision Current Source.  Cost: \$846. 

This model was chosen both for its precision and its compliance 
voltage.  Most diode laser drivers can't supply the 5V operating 
voltage of the gallium nitride laser.

\supplier{
MTI Corporation \\
5327 Jacuzzi St. Bldg. 3H \\
Richmond, CA 94804 \\
(510) 525-3070\\
http://www.mticrystal.com\\
\\
}
SO*10 x-cut 5mm$\times$5mm$\times$0.5mm quartz substrate, optically 
polished on both sides.  Cost: \$12.  

We glued this piece of quartz to a
metal support to mount in an optic holder.  

\supplier{
Nichia Corporation \\
Tokyo technical center \\
13F Tamachi Center Building, 34-7, Shiba 5-Chome \\
Minato-Ku, Tokyo 108-0014\\
Japan\\
{http://www.nichia.co.jp }\\
\\
}
NLHV500C: 5mW nominal wavelength 405 nm InGaN laser diode, 1000 hour
version.  Cost: \$1000. 

We tested two diodes, one with a wavelength of 400 nm, the
other 406 nm, with very similar results.  As InGaN laser diodes are a 
very young technology, it is reasonable to expect improvements in cost
and lifetime in the near future.

\supplier{
Perkin-Elmer Optoelectronics\\
22001 Dumberry Road \\
Vaudreuil Quebec J7V 8P7\\
CANADA\\
(450)-424-3300\\
http://opto.perkinelmer.com\\
\\
}
SPCM-AQR-13 Silicon APD Based Single Photon Counting Modules $\times$2.
Total cost: \$7200.

The -13 suffix describes the dark count rate.  This has very little
effect upon the coincidence count rate in the experiment, and any
suffix model can be used.  Note that the lead time for these detectors
can be several months.  Detectors require a +5 Volt power supply, not
included.

\supplier{
Thorlabs, Inc.\\
435 Route 206\\
P.O. Box 366\\
Newton, NJ 07860\\
(973) 579-7227\\
http://www.thorlabs.com\\
\\
}
C230TM-A: 0.55 NA aspheric lens, AR coated 350-600nm.
Collimating tube for 5.6mm laser diode and C230TM-A.
S7060: Laser diode socket for 5.6mm laser.
VC1 V-Clamp
SPW301: 3/8'' Spanner wrench.
SPW602: SM1 Series Spanner wrench.
CP02: 30mm Threaded cage plate $\times$2.
HPT1: 30mm cage X-Y translator $\times$2.
ER4: Extension rod, 4'' long $\times$8.
SM1V10: 1'' adjustable focus tube for 1'' optics $\times$4.
SM1V05: 1/2'' adjustable focus tube for 1'' optics $\times$2.
SM1T2: SM1 coupler, external threads $\times$2.
RSP1: Rotation stage for 1'' optics $\times$3.
Total cost: \$1040.

\supplier{
U-oplaz Technologies \\
21828 Lassen St, \#D\\
Chatsworth, CA 91311\\
(818) 678-1999\\
http://www.u-oplaz.com\\
\\
}
Two identical $\beta$-barium borate (BBO) crystals,
5mm$\times$5mm$\times$0.1mm with a crystal cut of 29\degree~, 
protective ``P-coating.''  Crystals custom-mounted face-to-face 
with one crystal rotated by 90\degree~ in a standard mount.
Cost: \$1400.

\bibliographystyle{revtex}

\end{document}